\documentstyle[12pt]{article}

\newcommand{\be}{\begin{equation}}
\newcommand{\ee}{\end{equation}}
\newcommand{\sk}{\stackrel{\rightarrow}{k}}
\newcommand{\sr}{\stackrel{\rightarrow}{r}}
\newcommand{\lgl}{\langle}
\newcommand{\rgl}{\rangle}
\newcommand{\dgr}{\dagger}
\newcommand{\om}{\omega}
\newcommand{\dlt}{\delta}
\newcommand{\prt}{\partial}
\newcommand{\ep}{\varepsilon}

\begin{document}

\begin{center}

{\Large{\bf Conditions for Nuclear--Matter Lasers} \\ [5mm]
V.I. Yukalov} \\ [3mm]

{\it Centre for Interdisciplinary Studies in Chemical Physics \\
University of Western Ontario, London, Ontario N6A 3K7, Canada \\
and \\
Bogolubov Laboratory of Theoretical Physics \\
Joint Institute for Nuclear Research, Dubna 141980, Russia}

\end{center}

\vspace{2cm}

\begin{abstract}

Conditions are analysed when in dense and hot nuclear matter large amounts
of Bose particles can be created. An intensive production of Bose particles 
is the main necessary condition for realizing their coherent emission
similar to radiation from photon lasers. The consideration is based on 
the multichannel model of nuclear matter. Analysis shows that possible
candidates for nuclear--matter lasing are mesons (mainly pions), dibaryons, 
and gluons.

\end{abstract}

\section{Introduction}

Since any kind of Bose particles shares the same statistical properties 
as photons, it sounds reasonable to pose a question as to whether it could
be feasible to realize coherent emission of other Bose particles by analogy
with the laser radiation of photons. This question is now intensively
discussed in connection with the possibility of realizing atom lasers
[1--7]. Experiments [8,9] show that Bose condensed atoms in a trap are in
a coherent state. Therefore a condensate released from the trap
propagates according to a single--mode wave equation represented by the
nonlinear Schr\"odinger equation [10--12]. Output couplers for Bose
condensed atoms are realized by means of short radiofrequency pulses
transferring atoms from a trapped state to a untrapped state [8,9,13,14].
With the help of additional external fields one could create Bose
condensates in non--ground states [15] or in vortex states [16], thus,
forming various modes of atom lasers.

Another possibility is related to the creation of a large number of pions
in hadronic, nuclear, and heavy--ion collisions [17]. In such collisions
up to hundreds of pions can be created simultaneously. When the density of
pions produced in the course of these collisions is such that their mean
particle separation approaches the thermal wave--length then
multi--particle interference becomes important. Strong correlations
between pions can result in the formation of a coherent state and in the
feasibility of getting a pion laser [18,19].

Coherent states are usually associated with Bose condensed states.
Therefore those particles that could exhibit Bose condensation under
extreme conditions characteristic of fireballs produced in heavy--ion
collisions could be also considered as candidates for lasing. For example,
such candidates could be dibaryons that, as was shown [20--23], can form a
Bose--Einstein condensate.

One of the main stipulations for the creation of coherent states is, as
is mentioned above, sufficient density of generated Bose particles. It is,
hence, necessary to understand what are the optimal conditions providing
the maximal possible density of bosons. It is the aim of this paper to
analyse the behaviour of dense and hot nuclear matter in order to answer
the questions -- what kind of bosons and under what proviso can be
generated in large quantities in such a matter.

\section{Multichannel Model}

To consider dense and hot nuclear matter, in which various kinds of
particles can be generated, we use the multichannel approach to clustering
matter [20,21]. The idea of this approach goes back to the methods of
describing composite particles [24--28]. The most complete basis for this 
problem was formulated by Weinberg [29--33]. According to this approach,
it is possible to introduce into any theory fictitious elementary particles,
or quasiparticles, without changing any physical predictions. To accomplish
this, the interaction among the original, truly elementary, particles must
be modified in the appropriate way. By "composite particles" one can mean
bound states or resonances. If fictitious elementary particles, 
quasiparticles, are introduced to take the place of all composite
particles, then perturbation theory can always be used. The modification
of the Hamiltonian weakens the original interaction enough to remove
divergencies. If such quasiparticles are introduced for each resonance or
bound state, then two--body scattering problems can always be solved by
perturbation theory. A nice account of the quasiparticle approach was
given by Weinberg in Ref.[34]. A resum\'e of this approach can be
formulated as follows: One introduces fictitious elementary particles into
the theory, in rough correspondence with the bound states of the theory.
In order not to change the physics, one must at the same time change the
potential. Since the bound states of the original theory are now
introduced as elementary particles, the modified potential must not
produce them also as bound states. Hence, the modified potential is
weaker, and can in fact be weak enough to allow the use of perturbation
theory.

Composite particles in other words are called clusters. Following the
multichannel approach to describing clustering matter [20,21], let us
consider an ensemble of particles that can form different bound states
interpreted as composite particles or clusters. A space ${\cal H}_i$ of
quantum states associated with a cluster of $z_i$ particles is termed an
$i$--channel. The number $z_i$ of particles forming a bound cluster is a
compositeness number. The average density of matter is a sum
\be
\rho=\sum_i z_i\rho_i\; ,
\ee
in which
\be
\rho_i =\frac{\zeta_i}{(2\pi)^3} \int n_i(\sk)d\sk
\ee
is an average density of $i$--channel clusters, $\zeta_i$ being a
degeneracy factor, and
$$
n_i(\sk) = \lgl a_i^\dgr(\sk) a_i(\sk)\rgl
$$
is a momentum distribution of the $i$--channel clusters. The statistical
weight of each channel is characterized by the channel probability
\be
w_i \equiv z_i \frac{\rho_i}{\rho}\; .
\ee

The Hamiltonian of clustering matter reads
\be
H =\sum_i H_i + CV\; ,
\ee
where $H_i$ is an $i$--channel Hamiltonian and $CV$ is a nonoperator term
providing the validity of the principle of statistical correctness [20,21],
$V$ being the system volume. Since strong short--range interactions
between original particles are included into the definition of bound
clusters, the left long--range interactions can be treated as week
[29--34]. These long--range interactions permit us to apply the
mean--field approximation resulting in an $i$--channel Hamiltonian
\be
H_i =\sum_k \om_i(\sk) a_i^\dgr(\sk) a_i(\sk)
\ee
with an effective spectrum
\be
\om(\sk) =\sqrt{k^2 + m_i^2} + U_i -\mu_i\; ,
\ee
where $m_i$ is an $i$--cluster mass; $U_i$, a mean field; and $\mu_i$ the
chemical potential of $i$--clusters. Then the momentum distribution of
$i$--clusters, in the Hartree approximation, takes the form
\be
n_i(\sk) =\frac{1}{\exp\{ \beta\om_i(\sk) \mp 1\} }\; ,
\ee
in which $\beta$ is inverse temperature; the upper or lower signs in (7)
stand for Bose-- or Fermi clusters, respectively. When the average baryon
density
\be
n_B = \sum_i \rho_i B_i\; ,
\ee
where $B_i$ is the baryon number of an $i$--cluster, is fixed, then the
chemical potentials of $i$--clusters,
\be
\mu_i =\mu_B B_i \qquad (n_B = const) \; ,
\ee are expressed through the baryon potential $\mu_B$ defined from (8).

The mean density of matter (1) may be written as the sum
\be
\rho =\rho_1 +\rho_z , \qquad \rho_1 \equiv \sum_{\{ i\}_1} \rho_i ,
\qquad \rho_z\equiv \sum_{\{ i\}_z} z_i\rho_i
\ee
of the density of unbound particles, $\rho_1$, and the density of particles 
bound in clusters, $\rho_z$. Then the conditions of statistical correctness
[20,21] are
\be
\left\lgl\frac{\dlt H}{\dlt\rho}\right\rgl = 0 \; , \qquad
\left\lgl\frac{\dlt H}{\dlt\rho_z}\right\rgl = 0 \; .
\ee

The original unbound particles in nuclear matter are quarks and gluons.
Their collection is named quark--gluon plasma. The mean--field potential
of the quark--gluon plasma can be written [20] as
\be
U_1 \equiv U(\rho) = J^{1+\nu}\rho^{-\nu/3}\; ,
\ee
where $J$ is an effective intensity of interactions and $\nu$ is an
exponent of a confining potential, $0<\nu\leq 2$. In what follows we take
$\nu\approx 2$. The mean field for $i$--channel clusters [20] reads
\be
U_i = z_i\left [ \Phi\rho_z + U(\rho) - U(\rho_z)\right ] \; ,
\ee
where $\Phi$ is a reference interaction parameter. With the potential
(12), we have
$$
U_i = z_i\Phi\rho_z + z_i J^{1+\nu}\left ( \rho^{-\nu/3} - \rho_z^{-\nu/3}
\right ) \; .
$$
From here and the condition of statistical correctness (11), we find the
correcting term
\be
C =\frac{\nu}{3-\nu} J^{1+\nu}\left ( \rho^{1-\nu/3} - \rho_z^{1-\nu/3}
\right ) -\frac{1}{2}\Phi \rho_z^2\; .
\ee

We have yet two undefined parameters, $J$ and $\Phi$. The first of them is
an effective intensity of interactions in the quark--gluon plasma, which
we take [20] as $J=225\; MeV$. The second, that is the reference parameter
$\Phi$, may be scaled [20,21] by means of nucleon--nucleon interactions
$V_{33}(r)$ as follows:
\be
\Phi =\frac{1}{9}\int V_{33}(r) d\sr\; .
\ee
Accepting for $V_{33}(r)$ the Bonn potential [35], we get $\Phi=35\; MeV\;
fm^3$. For nuclear matter of the normal baryon density $n_{0B}=0.167\;
fm^{-3}$, this gives an average interaction energy $\Phi n_{0B}=5.845\;
MeV$.

In this way, the model is completely defined and we can calculate all its
thermodynamic characteristics. For the pressure we have
\be
p =\sum_i p_i - C\; , \qquad p_i =\pm T\frac{\zeta_i}{(2\pi)^3}
\int\ln\left [ 1 \pm n_i(\sk)\right ] d\sk\; .
\ee
The energy density is
\be
\ep =\sum_i \ep_i + C \; , \qquad
\ep =\frac{\zeta_i}{(2\pi)^3}\int\sqrt{k^2 + m_i^2}\; n_i(\sk) d\sk
+ \rho_i U_i\; .
\ee
From here, we may find the specific heat and the reduced specific heat,
\be
C_V =\frac{\prt\ep}{\prt T}\; , \qquad
\sigma_V =\frac{T}{\ep} C_V\; ,
\ee
respectively, and the compression modulus
\be
\kappa^{-1}_T = n_B\frac{\prt p}{\prt n_B}\; .
\ee
One may also define an effective sound velocity, $c_{eff}$, by the ratio
\be
c_{eff}^2 = \frac{p}{\ep}\; .
\ee

Statistical weights of the corresponding channels are given by the channel
probabilities defined in (3). For the following analysis, it is convenient
to introduce also the plasma--channel probability
\be
w_{pl} =\frac{1}{\rho}\left ( \rho_g + \rho_u + \rho_{\bar u} +
\rho_d + \rho_{\bar d}\right )\; ,
\ee
where $\rho_g$ is the density of gluons, while other terms are the
densities of $u$-- and $d$--quarks and antiquarks, respectively. The
pion--channel probability
\be
w_\pi =\frac{2}{\rho} \left ( \rho_{\pi^+} + \rho_{\pi^-} +
\rho_{\pi^0}\right )
\ee
is expressed through the densities of $\pi^+,\;\pi^-$, and $\pi^0$ mesons.
The probability of other meson channels, except pions, is
\be
w_{\eta\rho\om} = \frac{2}{\rho}\left ( \rho_\eta + \rho_{\rho^+} +
\rho_{\rho^-} + \rho_{\rho^0} + \rho_\om \right ) \; .
\ee
The nucleon--channel probability writes
\be
w_3 = \frac{3}{\rho} \left ( \rho_n +\rho_{\bar n} +
\rho_p + \rho_{\bar p} \right )
\ee
containing the densities of neutrons, antineutrons, protons, and
antiprotons. We calculate also the probabilities of multiquark channels,
such as the dibaryon--channel probability
\be
w_6 = \frac{6}{\rho}\left ( \rho_{6q} + \rho_{6\bar q} \right ) \; ,
\ee
and, analogously, the $9$--quark and $12$--quark channel probabilities.

Now we can analyse the thermodynamic behaviour of the described model in
order to define what kinds of Bose particles and under what conditions can
be generated in large quantities, that is, when the corresponding
Bose--channel probabilities are maximal. The choice of parameters is the
same as in Ref. [20].

\section{Analysis}

The multichannel model of nuclear matter described in the previous section
has been solved numerically. The pressure (16) is shown in Fig.1 as a
function of temperature $\Theta=k_BT$ in $MeV$ and of relative baryon
density $n_B/n_{0B}$. The pressure is a monotonic function of its
variables as well as the energy density (17) in Fig.2. But it is
interesting that their ratio (20) in Fig.3 is a nonmonotonic function
displaying a maximum at temperature around $T_d=160\; MeV$. The latter, as
will be clear from the following, can be associated with the temperature
of the deconfinement crossover. The specific heat and the reduced specific
heat given in (18) are presented in Fig.4 and 5, respectively. The
compression modulus (19) is shown in Fig.6. Again, the maxima of the
reduced specific heat and the compression modulus can be associated with
the deconfinement crossover. The following Figs. 7 to 11 present the
behaviour of the channel probabilities for the quark--gluon plasma (21),
pions (22), other mesons (23), nucleons (24) and dibaryons (25). Since the
possibility of the appearance of the dibaryon Bose condensate is of
special interest, we show in Fig. 12 the corresponding channel probability
$w$. The Bose condensates of heavier multiquark clusters do not arise. The
channel probabilities of such heavier clusters are negligibly small being,
for instance, for $9$-- and $12$--quark clusters less than $10^{-3}$ and
$10^{-5}$, respectively. We show also in Figs. 13 to 15 the channel
probabilities, as functions of the relative baryon density $n_B/n_{0B}$ at
zero temperature, for the quark--gluon plasma (21), nucleons (24), and
dibaryons (25).

The analysis demonstrates that the maximal density of pions can be
generated around the temperature $T\approx 160\; MeV$ of the deconfinement
crossover at low baryon density $n_B < n_{0B}$. The corresponding channel
probability of pion production can reach $w_\pi\approx 0.6$. The total
probability of other meson channels reaches only $w_{\eta\rho\om}\approx
0.16$ at $T\approx 200\; MeV$ and $n_B < n_{0B}$. However, the generation
of these mesons is more noticeable than that of pions at high temperatures
and baryon densities, although being always not intensive, with the
related probability not exceeding the order of $10^{-1}$.

The optimal region for the creation of dibaryons, where their channel
probability reaches $w_6\approx 0.7$, is the region of low temperatures
$T<20\; MeV$ and the diapason of baryon densities $n_B/n_{0B}\approx
5-20$. At zero temperature their probability rather slowly diminishes
with increasing the baryon density, so that at $n_B\approx 100n_{0B}$, we
have $w_6\approx 0.4$. At low temperatures dibaryon form a Bose--condensed
state.

Above the deconfinement crossover temperature $T_d\approx 160\; MeV$,
there is an intensive generation of gluons in the quark--gluon plasma. At
sufficiently high temperatures, gluon radiation can, in principle, become
so intensive that to acquire a noticeable coherent component.

Thus, the most probable candidates for realizing laser generation are
pions, dibaryons, and gluons. Each kind of these Bose particles has its
own region where the corresponding channel probability is maximal. For
pions it is $T\approx 160\; MeV$ and $n_B< n_{0B}$; for dibaryons, $T<20\;
MeV$ and $n_B\approx (5-20)n_{0B}$; and for gluons, this is the
high--temperature region $T>160\; MeV$. If it is feasible to realize the
corresponding conditions, one could get a pion laser, dibaryon laser, or
gluon laser, respectively. Note that to realize such a lasing in reality
one has to accomplish several other requirements of which we are
considering here only one necessary condition.

It is also worth noting that if one tries to achieve the desired
conditions of lasing in the process of hadronic or heavy--ion collisions
then one can get only a pulsed radiation of Bose particles. If the
lifetime of a fireball formed during a collision is longer than the
local--equilibrium time then the quasiequilibrium picture of the process
is permissible. In such a case, it is possible to use the multichannel
model, as is described here, with temperature and baryon density given 
as functions of time, the time dependence being in accordance with the
related fireball expansion.

\vspace{5mm}

{\bf Acknowledgement}

\vspace{2mm}

I am grateful to E.P. Yukalova for useful discussions. A grant from the
University of Western Ontario, London, Canada, is appreciated.

\newpage

\begin{center}
{\bf Figure captions}
\end{center}

{\bf Fig.1.} The pressure (in units of $J^4$) of the multichannel model.

\vspace{5mm}

{\bf Fig.2.} The energy density (in unites of $J^4$) on the
temperature--baryon density plane.

\vspace{5mm}

{\bf Fig.3.} The pressure--to--energy density ratio related to an
effective sound velocity squared, $c^2_{eff}$.

\vspace{5mm}

{\bf Fig.4.} The specific heat (in units of $J^3$) for the multichannel
model.

\vspace{5mm}

{\bf Fig.5.} The reduced heat displays a maximum that can be associated
with the deconfinement crossover.

\vspace{5mm}

{\bf Fig.6.} The compression modulus (in units of $J^4$) for the
multichannel model.

\vspace{5mm}

{\bf Fig.7.} The channel probability of the quark--gluon plasma.

\vspace{5mm}

{\bf Fig.8.} The pion channel probability.

\vspace{5mm}

{\bf Fig.9.} The total probability of other, except pion, meson channels.

\vspace{5mm}

{\bf Fig.10.} The nucleon channel probability.

\vspace{5mm}

{\bf Fig.11.} The dibaryon channel probability.

\vspace{5mm}

{\bf Fig.12.} The channel probability of Bose--condensed dibaryons.

\vspace{5mm}

{\bf Fig.13.} The plasma channel probability at zero temperature as a
function of the relative baryon density.

\vspace{5mm}

{\bf Fig.14.} The nucleon channel probability at zero temperature.

\vspace{5mm}

{\bf Fig.15.} The dibaryon channel probability at zero temperature.

\end{document}